\documentclass[aip,apl,reprint,numerical,longbibliography,superscriptaddress,amsfonts,amsmath,amssymb]{revtex4-1}
\usepackage[utf8]{inputenc}
\usepackage{graphicx}

\usepackage{booktabs}
\usepackage{threeparttable}
\usepackage{url}
\usepackage{siunitx}
\begin{document}
\title{\Large{Crowdsourcing accurately and robustly predicts Supreme Court decisions}}


\affiliation{Illinois Tech - Chicago Kent College of Law}
\affiliation{CodeX - The Stanford Center for Legal Informatics}
\affiliation{South Texas College of Law  Houston}
\affiliation{LexPredict, LLC}

\author{\large{Daniel Martin Katz}}
\thanks{Corresponding Author: \\Daniel Martin Katz, dkatz3@kentlaw.iit.edu}
\affiliation{Illinois Tech - Chicago Kent College of Law}
\affiliation{CodeX - The Stanford Center for Legal Informatics}
\affiliation{LexPredict, LLC}
\author{\large{Michael J. Bommarito II}}
\affiliation{Illinois Tech - Chicago Kent College of Law}
\affiliation{CodeX - The Stanford Center for Legal Informatics}
\affiliation{LexPredict, LLC}
\author{\large{Josh Blackman}\vspace{5mm}}
\affiliation{South Texas College of Law  Houston}
\affiliation{LexPredict, LLC}

\date{\today}

\begin{abstract}
	\hrulefill\\
	Scholars have increasingly investigated ``crowdsourcing'' as an alternative to expert-based judgment or purely data-driven approaches to predicting the future.  Under certain conditions, scholars have found that crowdsourcing can outperform these other approaches.  However, despite interest in the topic and a series of successful use cases, relatively few studies have applied empirical model thinking to evaluate the accuracy and robustness of crowdsourcing in real-world contexts. In this paper, we offer three novel contributions.  First, we explore a dataset of over 600,000 predictions from over 7,000 participants in a multi-year tournament to predict the decisions of the Supreme Court of the United States.  Second, we develop a comprehensive crowd construction framework that allows for the formal description and application of crowdsourcing to real-world data.  Third, we apply this framework to our data to construct more than 275,000 crowd models.  We find that in out-of-sample historical simulations, crowdsourcing robustly outperforms the commonly-accepted null model, yielding the highest-known performance for this context at 80.8\% case level accuracy.  To our knowledge, this dataset and analysis represent one of the largest explorations of recurring human prediction to date, and our results provide additional empirical support for the use of crowdsourcing as a prediction method.
\end{abstract}

\maketitle
\section{Introduction}
\label{section:introduction}
The future appears to be intrinsically uncertain despite our best attempts to divine the mechanisms of a clockwork universe.  Yet even though we are not able to fully understand the precise mechanics underlying many phenomena, we are still able to predict the future in many circumstances.  We can assess future medical risk, \cite{weng2017can}$^{,}$\cite{obermeyer2016predicting} forecast the weather,\cite{bauer2015quiet} trade securities,\cite{kazem2013support} predict future sales,\cite{wu2015future}$^{,}$\cite{kim2015box}  and identify future geopolitical events.\cite{mandel2014accuracy}$^{,}$\cite{tetlock2014judging}  Indeed, individuals and organizations, both large and small, spend a substantial amount of effort and energy generating various types of forecasts.  In some cases, predictions appear as if they perfectly anticipate the future, but in many other cases, the predictions of even our best experts and models appear to be no better than random.\cite{tetlock2006expert}$^{,}$\cite{spann2009sports}$^{,}$
\cite{shepperd2012evaluating}  

While predictions vary in many ways, they all share one characteristic - they are the product of at least one of three sources: (i) a single human, typically an \textit{expert}, (ii) a group of humans, commonly referred to as a \textit{crowd}, or (iii) a model - statistical, rule-based, or otherwise - i.e., an \textit{algorithm}.  It is also possible to construct a ``model of models'' or \textit{ensemble} of all three approaches; in fact, crowds are themselves a ``model of models'' built on predictions of individuals.\cite{lee2015Crowd}$^{,}$\cite{jordan1994hierarchical} Throughout human history, and long before the creation of the scientific method, humans have been using all three of these approaches, including such model-based predictions as augury and astrology.  However, in recent years, attention has increasingly focused  on ``scientific'' approaches to prediction.

In this paper, we investigate the prediction of future events using crowds and models thereof.  This practice, commonly referred to as ``crowdsourcing,'' has received increasing interest over the past years across a wide variety of problems.\cite{atkinson2007did}$^{,}$\cite{glaeser2016predictive}$^{,}$\cite{pickard2011time}$^{,}$\cite{park2017crowdsourcing}  Despite interest in the topic and a series of successful use cases, relatively few studies have rigorously applied empirical model thinking to evaluate both the accuracy and robustness of crowdsourcing in real-world contexts. This paper contributes to the extant literature on prediction by presenting results from an online prediction tournament called ``FantasySCOTUS.'' \cite{blackman2011fantasyscotus}  Over the past several years, participants with diverse backgrounds and experience levels have competed to predict the decisions of the Supreme Court of the United States (SCOTUS).  Collectively, more than 7,000 participants have cast over 600,000 predictions across nearly 450 cases between 2011 and 2017.  Using this data, we are able to empirically evaluate the performance of a model space of more than 275,000 different crowdsourcing models. 

We begin in Section \ref{section:data} by describing the FantasySCOTUS dataset, including the history of the tournament and basic facts about the dataset.  Next, in Section \ref{section:research_goals_and_prior_work}, we describe in detail our research goals and motivating principles, and how this research sits in the context of prior work on crowdsourcing,  prediction and legal prediction.  Once these goals and principles are outlined, we proceed in Section \ref{section:crowd_construction_framework} to  formalize a crowd construction framework that can be applied generally to any crowdsourcing problem.  We apply this framework in Section \ref{section:model_testing_results} to evaluate the performance of over 275,000 crowd models across a range of crowd construction parameters.  We find that crowdsourcing yields both accurate and robust results, including a peak performance of over 80\% accuracy, which is the highest documented performance for Supreme Court prediction.  We conclude in Section \ref{section:conclusion_future_work} by summarizing our findings and discussing future research on the topic.

\section{Data}
\label{section:data}
\subsection{FantasySCOTUS Background}
FantasySCOTUS is an online prediction tournament in which participants cast predictions about ``SCOTUS'' - the Supreme Court of the United States. In the fashion of a fantasy sports league, correct predictions for the votes of each Justice of the Supreme Court are rewarded with points, and these points are tallied through the Term of the Supreme Court, beginning each October and ending in the following summer.\cite{blackman2011fantasyscotus}  Participation in FantasySCOTUS is open to individuals without any ``buy-in'' or qualification, although prizes are restricted to US citizens.  Whether an individual is young or old, expert or novice, they can register for free and predict.  Thus, while the set of participants certainly exhibits some selection biases, the crowd as a whole contains a wide range of participants with varied backgrounds competing for prizes.

Throughout the last eight terms - October 2009 through October 2016 - the contest rules and prize structure have varied, with first place prize ranging from a non-cash ``golden gavel'' trophy to a \$10,000 cash prize. Yet while rules and incentives for contest participants have varied, the rules of the Court have not.  Each year, \textit{petitioners} ask the Supreme Court to hear their case against an opposing party, called the \textit{respondent}.  If four or more of the Court's Justices agree to hear the case, then the Court generally accepts the case onto its docket by granting a petition for a writ of \textit{certiorari}.  Once \textit{certiorari} is granted, FantasySCOTUS generally lists these new cases on its site within days.  In the event that Justices announce their recusal from a case, they are marked as ineligible on the case; for all other Justices, FantasySCOTUS participants may begin predicting vote outcomes as soon as the cases are available on the site.

The petitioners and respondents then submit written materials supporting their positions, and hold one or more in-person arguments before the Court.  Following oral argument, the Court may issue an opinion on any Monday, Tuesday, or Wednesday around 10AM Eastern as designated on its calendar.  External observers do not know in advance which cases will be decided on which days, but in practice, even the simplest opinions may take the Court at least one month to decide, draft, and publish.  Participants therefore know that, once the Court has completed its request for materials and argument, they must cast their predictions prior to such designated announcement dates.

Ultimately, the votes of each participating Justice are made public in the published opinions of cases.  In general, they vote to either affirm or reverse the status quo represented by the lower court's prior decision.  While these rules do not apply in some cases, such as decisions of a split court or \textit{per curiam} decisions, all but a small percent of Supreme Court votes can be represented as simple binary outcomes.  Over the last decade, the Court has ruled on approximately 60-90 cases per term, resulting in approximately 500-800 binary Justice votes per term.

\subsection{FantasySCOTUS Data}
As with many sites, FantasySCOTUS has evolved since it was first launched in 2009,\cite{blackman2011fantasyscotus} migrating from its original blog format to its current purpose-built web application.  Beginning with the 2011 term, the underlying data model and contest rules have been relatively consistent, and we therefore begin the analysis from this term.  While data from prior terms is available, the 2009 and 2010 formats, which were focused on whether a Justice signed to the majority opinion, unfortunately preclude comparison with later terms.

Given this constraint, our period of analysis thus begins with the 2011 term and ends with the completion of the 2016 term, resulting in a total of six terms available for analysis.  For each case listed on FantasySCOTUS during these years, we also compare our cases against those listed in the Supreme Court Database (SCDB).\cite{spaeth2016}  In some situations, cases were listed on FantasySCOTUS but never officially decided by the Supreme Court and therefore unpublished in SCDB.  For example, cases such as \textit{UBS v. Unión de Empleados} and \textit{Cline v. Oklahoma Coalition for Reproductive Justice} received thousands of FantasySCOTUS predictions, but were subsequently settled by the parties or dismissed by the Court prior to formal decision.  In these cases, we omit such records from this analysis, as no measurable contest outcome occurred and these cases cannot be merged with SCDB data.

Once these records are handled and SCDB data is merged, the resulting dataset contains 7,284 unique participants, 425 listed cases, 182 dates of decision, and 636,859 predicted votes for 10 separate Justices, including Roberts, Scalia, Kennedy, Thomas, Ginsburg, Breyer, Alito, Sotomayor, Kagan, and Gorsuch.  In some situations, these cases may be re-argued in successive terms, resulting in some cases counting twice in Table \ref{table:table1}.  These predictions begin with \textit{Greene v. Fisher}, decided in November of 2011, and end with \textit{Davila v. Davis}, decided in June 2017.

This dataset is from the real world, for better \textit{and} worse, and there are many noteworthy caveats.  First, 636,859 predictions is many fewer than the theoretical upper bound - the number of participants times the number of cases and Justices, which would total approximately 30 million; in fact, it is less than 3\% of the total number of predictions possible under maximal conditions.  This sparsity is partially the result of Justice recusals or absences, such as the vacant seat left after Justice Scalia unexpectedly passed away in 2016.  More accurately, however, this sparsity is the first hint that our crowd changes over time.  Like many other real-world crowds, participants join and exit, either temporarily or permanently, resulting in high turnover and low stability period-over-period.  Within our window of analysis, only six participants participated in every term, and 108 participants participated in at least three terms.  The mean and median number of terms participated in are 1.09 and 1.00, respectively, demonstrating the high level of attrition bias. As demonstrated below, this changing crowd composition presents challenges for our analysis relative to theoretical models.

Second, while most participants do not predict most cases, some update their predictions in a single case multiple times.  The average number of predictions per participant per outcome is approximately 1.1, and some cast many more predictions; in fact, some of our highest-scoring participants act like ``foxes,'' frequently updating their predictions for some cases.\cite{tetlock2006expert}  While patterns of behavior like ``foxes'' and ``hedgehogs'' are worthy of additional research, for the purposes of this paper, we exclude from our analysis all ``non-final'' predictions.  The resulting set of ``final'' predictions - the last recorded prediction for each participant for each Justice in each case that was cast before midnight on the date of decision of the case - consists of 545,845 remaining predictions.

Third, while the number of active participants per term has fallen, the average engagement level - as measured by the average number of predicted cases per participant per term - has increased.  These variations are most likely due to changes in prize structure and marketing.  In the 2013, 2014, and 2015 terms, higher prizes and more marketing assistance for the FantasySCOTUS contest attracted more new participants.  Though this resulted in large numbers of new participants, these participants also dropped out faster than average, predicting fewer cases per term before leaving.  Thus, while the population of participants is lower in 2015 and 2016 than in 2013 and 2014, the remaining population's engagement rate is higher.  This trend and other per-term summary statistics are shown in Table \ref{table:table1} below. 

\setlength{\tabcolsep}{0.4em} 
{\renewcommand{\arraystretch}{1.1}%
	\begin{table}[htb]
		\footnotesize
		\begin{tabular}{|c|c|c|c|c|}
			\hline
			\textbf{Term} & \textbf{Predictions} & \textbf{Participants} & \textbf{Cases} & \textbf{Engagement} \\\hline
			2011          & 81,320               & 1,336                 & 82             & 6.9                 \\
			2012          & 76,786               & 1,348                 & 81             & 6.4                 \\
			2013          & 115,326              & 1,951                 & 77             & 6.7                 \\
			2014          & 127,138              & 1,824                 & 73             & 7.8                 \\
			2015          & 87,084               & 927                   & 71             & 11.5                \\
			2016          & 58,191               & 621                   & 65             & 11.7                \\
			\hline
		\end{tabular}
		\caption[Table caption text]{Summary statistics for FantasySCOTUS Participation data by term.}
		\label{table:table1}
	\end{table}
	
	In order to promote reproducible research and further study, participant-anonymized FantasySCOTUS data will be released to GitHub and the FantasySCOTUS website.\cite{fantasyscotus}  In addition, the source code used to produce the simulations and figures in this publication will be released there.  We encourage other researchers to build on our efforts in future research.
	
	\section{Research Principles and Prior Work}
	\label{section:research_goals_and_prior_work}
	In this section, we set out our research goals, including key questions and principles, and discuss prior work in this area.
	
	\subsection{Research Principles}
	\label{section:research_goals}
	In this paper, we investigate whether and under what conditions crowdsourcing can provide accurate predictions for Supreme Court decisions relative to other available methods.  Our approach to answering these questions is guided by three principles.  First, we must work with real data, even when it deviates far from theory.  Unlike much of the theoretical work in the field,\cite{condorcet1785essay}$^{,}$\cite{hong2004groups} most actual crowds do not have a fixed, odd number $N$ of participants with known probability distributions. FantasySCOTUS participants, by contrast, join and exit the crowd day by day and case by case.  Participants act, as humans generally do, in ways that we cannot control.  They possess a range of changing and heterogeneous individual attributes, including differing cognitive processes, information sets, and effort levels.  In order to translate theory to practice, models must support a number of dynamics, including the addition of new predictors with unknown behaviors, the non-stationarity of predictor behavior, and both the \textit{temporary} and \textit{permanent} removal of predictors.  The literature has confronted some of these challenges but many remain unresolved.\cite{de2014follow}$^{,}$\cite{creamer2010automated}$^{,}$\cite{erven2011adaptive}$^{,}$\cite{hung2013evaluation}$^{,}$\cite{devaine2013forecasting}$^{,}$\cite{vovk2009prediction}
	
	Second, in order for our results to accurately reflect what historical performance would have been, we must ensure that models produce only out-of-sample predictions - that is, models must \textit{not} incorporate any in-sample or ``future'' information that would not have been knowable as of the simulated date.  For example, we \textit{cannot} use information about a participant's accuracy or rank in February of 2015 to predict a case from January of 2015.
	
	Third, while we evaluate the performance of a few illustrative model configurations, we are not merely interested in the performance of \textit{a single crowd} per se, but of \textit{crowdsourcing generally}. In other words, we evaluate our primary claims for an entire \textit{space of models}, not just for a single model.  To do so, we must develop a framework for model construction, select a parameter space, generate many simulated historical crowds to analyze, and report the aggregate results. 
	
	We can summarize these three modeling principles as follows:
	\begin{enumerate}
		\item \textbf{Empirical}: Our analysis should use empirical data and directly confront properties that are often present in real-world crowds.
		\item \textbf{Out-of-sample}: Our analysis should reflect the historically-accurate information set available to predictors, allowing us to faithfully reproduce how a given model would have actually performed.
		\item \textbf{Model space, not merely a single model}: Our analysis and its findings should be based on the aggregate results of an entire space of models.
	\end{enumerate}
	
	\subsection{Prior Work}
	\label{section:prior_work}
	There is substantial prior work both on the science of crowdsourcing and on legal prediction, and it is important to understand how this paper contributes and compares to both of these extant literatures. 
	
	Given the importance of forecasting to many individuals and institutions, there is increasing interest in the science of prediction.\cite{subrahmanian2017predicting}$^{,}$\cite{kleinberg2015prediction}$^{,}$\cite{breiman2001statistical} Indeed, the field of predictive analytics is a fast-growing area of both industrial interest and academic study.  Although much of the forecasting literature has been published in fields such as statistics, computer science, and artificial intelligence, social science has recently begun to focus on this area as well. \cite{nay2017predicting}$^{,}$\cite{weidmann2010predicting}$^{,}$\cite{lazer2009computational}  In an effort to bring greater rigor to the challenge of forecasting economic, political, or social outcomes, and notwithstanding the fact that causal inference is clearly the dominant methodological orientation in the social sciences,\cite{breiman2001statistical}$^{,}$\cite{gelman2011causality}$^{,}$\cite{mullainathan2017machine} there have been a range of important contributions published over the past few years with implications for a variety of academic fields.\cite{hofman2017prediction}$^{,}$\cite{mellers2015identifying}$^{,}$\cite{gonzalez2008understanding}$^{,}$\cite{nofer2014crowds}$^{,}$\cite{de2002predicting}$^{,}$\cite{kleinberg2017}
	
	Of greatest relevance to the current inquiry is the growing number of papers focused on predicting the actions of legal institutions and actors.\cite{katz2017general}$^{,}$\cite{aletras2016predicting}$^{,}$\cite{mcshane2012predicting}$^{,}$\cite{ruger2004supreme}  Given the overall significance of legal decisions not only to individual participants but also to capital markets\cite{katz2015law} and society as a whole,\cite{epstein2001supreme}$^{,}$\cite{epstein1992supreme} the development of rigorous quantitative legal predictions\cite{katz2012quantitative} can have wide-spread benefits. In existing literature, scholars have typically applied well-known algorithms or variants thereof to forecast the decision-making processes of legal actors and institutions. \cite{katz2017general}$^{,}$\cite{ruger2004supreme} In addition, the literature has examined the performance of certain subject matter experts and compared their performance to statistical models.\cite{ruger2004supreme}$^{,}$\cite{grossman2010technology}  Algorithmic approaches have in some cases matched or exceeded the accuracy of subject matter experts.
	
	As highlighted in Section \ref{section:introduction}, crowdsourcing can provide an alternative approach to expert judgment and algorithmic decision-making.  There has been significant work to date on both theoretical \cite{condorcet1785essay}$^{,}$\cite{hong2004groups}$^{,}$\cite{hung2013evaluation} and empirical analysis of crowdsourcing.\cite{atkinson2007did}$^{,}$\cite{glaeser2016predictive}$^{,}$\cite{pickard2011time}$^{,}$\cite{park2017crowdsourcing}  Unfortunately, much of the existing work falls short relative to one or more of our research principles.  In particular, much of the literature either lacks real data or relies on the analysis of a single model selected \textit{post hoc}.  To our knowledge, no other study presents an analysis of crowdsourcing that evaluates the historical out-of-sample performance of a real \textit{model space} of crowds. 
	
	Further, across the literature, there is substantial conceptual confusion regarding what constitutes crowdsourcing and how crowds are constructed.\cite{estelles2012towards}  At their core, crowds are just the aggregation of individual predictive signals, each of which may or may not provide predictive insight.  The study of crowdsourcing is an exploration of how and when ``collective knowledge can be pooled together to address problems more efficiently and accurately than decisions from individuals.''\cite{blackman2011fantasyscotus}  While there is often a strong desire to seek wisdom from crowds, in reality, crowdsourcing is 
	as much a process of segmentation as aggregation.  Many crowdsourcing methods seek to boost predictive performance by identifying and over-weighting the best-performing players.  In general, however, crowd models vary widely in their construction, and crowdsourcing does not refer to a specific technique or algorithm.  In Section \ref{section:crowd_construction_framework} below, we seek to address this gap in the literature by developing a formal crowd construction framework.

	\section{Crowd Construction Framework}
	\label{section:crowd_construction_framework}
	As discussed in Section \ref{section:research_goals} above, we want to evaluate not only the performance a of specific crowdsourcing model but also the overall space of crowd models.  To do so, we first define a crowd construction framework which features varying potential hyperparameters and configurations.  
	\subsection{Model Notation}
	Let $d$ be an index for Supreme Court case or ``docket'' and let $T(d)$ be the date of decision for docket $d$.  Let $N(d)$ be the number of participants with at least one prediction for a case $d$, and the set of such participants be $S(d)$.  Then let the crowd vector for a Justice $j$ in case $d$ be $\vec{C^j}(d, S(d)) = \left(c^j_1(d), \cdots, c^j_{N(d)}(d)\right)$, where $c^j_i(d)$ is the prediction of the $i^{th}$ participant for the $j^{th}$ Justice in the $d^{th}$ case.  $c^j_i(d)$ is $1$ when participant $i$ predicts that Justice $j$ will vote to reverse in case $d$, else $0$.  Let the actual observed vote of Justice $i$ in case $d$ be $v^j(d)$.  Finally, let the information state $\mathcal{I}(t)$ be the set of $C^j(d)$ and $v^j(d)$ for all $d$ s.t. $T(d) < t$ - that is, let the information state reflect knowable participant predictions and Justice votes prior to the date of decision.
	
	A crowd model is a function $M^j(d)$ that takes as input the information state $\mathcal{I}(t)$ and crowd vector $\vec{C^j}(d, S(d))$ and yields a score in $[0, 1]$, where $0$ implies affirm and $1$ implies reverse for Justice $j$.  Some models, but not all, can be represented as linear transforms of $\vec{C^j}(d)$, as we shall explore below.
	
	\subsection{Crowd Subsets}
	\label{section:crowd_subsets}
	In some cases, such as in Condorcet juries, models may use predictions from all participants.  In other cases, however, models may only take into account information from some subset of participants.  These subsets can be defined either positively, through inclusion rules, or negatively, through exclusion rules.  Let a crowd subset $S^*(d) \subseteq S(d)$ be defined as the set $\{i : s(i, \mathcal{I}(T(d)) = 1 \ \forall i \in S(d)\}$, where $s(\cdot)$ encodes for one or more inclusion rules.  Common inclusion rules for defining subsets are listed below:
	
	\begin{enumerate}
		\item \textbf{Experience}: Let $\text{XP}(i, \mathcal{I}(d))$ be the number of cases that participant $i$ has predicted as of case $d$.  Then an experience threshold function $s(i, \mathcal{I}(d)) = \mathbb{I}_{\text{XP}(i, \mathcal{I}(d)) \geq p}$. In other words, a participant $i$ must have predicted at least $p$ cases.
		      
		\item \textbf{Performance}: Let $P(i, \mathcal{I}(d))$ be a performance function such as accuracy or F1 score.  Then a performance threshold function $s(i, \mathcal{I}(d)) = \mathbb{I}_{P(i, \mathcal{I}(d)) \geq a}$.  In other words, a participant $i$ must have a Justice or case-level accuracy or F1 score of at least $a$\%.  Performance can be defined using either broad or specific measures; for example, participants might be ranked based on their accuracy over all time and for all Justices, or they might be ranked only on their accuracy in the last 20 cases for a particular Justice.
		      
		\item \textbf{Rank}: Let $R_P(i, \mathcal{I}(d))$ be the rank of a participant $i$ under some performance measure $P$. Then a rank threshold function $s(i, \mathcal{I}(d)) = \mathbb{I}_{R_P(i, \mathcal{I}(d)) \leq r}$.  In other words, a participant $i$ must be in the top $r$ participants as ranked by some performance measure.  The decision theoretic strategy commonly known as ``follow the leader'' (FTL) can be seen as a rank threshold function with $r = 1$ or $r=0$ for ranks beginning with 1 or 0, respectively.  Furthermore, setting the rank threshold can be seen as analogous to choosing a ``crowd size'' in other literature.
		      
		\item \textbf{Statistical}: Let $F(i, \mathcal{I}(d))$ be the $p$-value or power of a statistical test.  Then a statistical threshold function $s(i, \mathcal{I}(d)) = \mathbb{I}_{F(i, \mathcal{I}(d)) \leq p^*}$.   In other words, a participant $i$ must pass some statistical test $F$, such as a binomial or t-test, with some given significance or power.
	\end{enumerate}
	
	These subset rules can be applied either independently or in combination.  For example, a crowd subset can be defined by combining experience and rank rules to create a crowd of the top 10 participants who have predicted at least 10 cases, or the top 25 participants who pass a one-sided binomial test with a $p$-value less than $0.05$.
	
	\subsection{Linear Crowd Models}
	Many crowd models can be represented as linear transforms of a crowd subset, and for such models, weight vectors are the key to understanding behavior.  Most research on crowds has, until recently, relied on this approach.  Below, we outline common approaches for defining weight vectors.
	
	\begin{enumerate}
		\item \textbf{Equal Weight}: Equal weight models, as the name implies, assign an equal weight to the predictions of all participants in a crowd subset - i.e., for a crowd subset $S^*(d)$, then an equal weight model $M^j(d)$ would be defined as $\frac{1}{|S^*(d)|} \sum_{i \in S^*(d)} c^j_i(d)$.  Such models are equivalent to majority-rule voting frameworks like Condorcet juries, and therefore the predictions of all participants receive equal weight in the final model prediction.
		      
		\item \textbf{Linear Weight}: Linear weight models can be used in combination with a performance or experience function or ranking, such as $R_P$ above.  A linear weight model $M^j(d)$ would be defined as $\frac{1}{\sum_{i \in S^*(d)} R_P(i)} \sum_{i \in S^*(d)} R_P(i) \cdot c^j_i(d)$.  In such models, for example, a participant who has predicted twice as many cases as another might have his or her predictions receive twice as much weight.
		      
		\item \textbf{Exponential Weight}: Exponential weight models are often used in combination with a performance or experience ranking like $R_P$ above.  An exponential weight model $M^j(d)$ would be defined as $\frac{1}{\sum_{i \in S^*(d)} \exp(-\alpha \cdot R_P(i))} \sum_{i \in S^*(d)} \exp(-\alpha \cdot R_P(i)) \cdot c^j_i(d)$ for some scale factor $\alpha$.  If the rank function $R_P(i)$ is defined such that no ties are produced, then the finite series sums above can also be further simplified.  The parameter $\alpha$ can also be used to produce widely varying behaviors.  For example, when $\alpha=0$, then we recover the equal weight model above, but for sufficiently large $\alpha$, we recover the ``follow-the-leader'' model.
	\end{enumerate}
	
	\subsection{``Machine Learning'' Crowd Models}
	\label{section:machine_learning_models}
	Not all models can be represented as weight vectors applied to a crowd subset through dot product.  For example, a more general class of ``machine learning'' models that defy this particular characterization (but might otherwise be effective) include tree-based models, neural network models, or logistic or support vector models.  Such models still produce a result in $[0, 1]$ -  through classification, probability scoring, or regression and calibration - but do so in ways that may involve non-linear or discontinuous transforms.
	
	More importantly, such models may incorporate additional information or features more naturally than simple linear transformations.  For example, suppose that we would like to augment our information state $\mathcal{I}$ with information about the identity of the petitioners and respondents or the nature of specific legal questions involved in a given case.  It is unclear how such information could be incorporated in a principled and statistical manner into an equal-weight (``majority-rule'') voting model.  While such models explain how to weight the predictions of \textit{participants}, they do not provide any guidance on what to do with \textit{non-participant} information.  If, instead, a model such as a random forest or support vector machine is being used, then we can simply encode this non-participant information as additional features for the model to consider.  As with any machine learning model, we allow the model to determine the relative feature weights or coefficients based on loss or error rates in the training sample.  While there is reason to believe that such machine learning models can provide additional benefits over linear models, in this paper,  we focus on crowdsourcing generally and therefore exclude models that include non-participant information.
	
	\section{Model Testing and Results}
	\label{section:model_testing_results}
	Imagine if the following story were real.  In October 2011, a million-dollar prize was announced to predict six years of decisions of the Supreme Court of the United States using crowdsourcing data.  Thousands of teams from across industry and academia respond to the challenge, each making slightly different but reasonable choices about how to cast their predictions.  Some teams develop models that rely more on ``following the leader(s),'' while other teams listen for the ``wisdom of the crowd'' at large.  Some teams develop complicated weighting schemes, while others stick with simple equal weighting.  Some teams even update their models over the six year challenge, iterating from their first attempt to explore more and more of the model space.  By the time that the tournament comes to an end, tens of thousands of different models have been tested across a wide range of choices.  Overall, how would these teams and their models have performed?  
	
	Our goal in this section is to answer this question by plausibly reconstructing the imagined tournament using the framework developed in Section IV above.  We exhaustively simulate models over a wide range of reasonable choices, and then analyze the performance of these models in aggregate.  In this analysis, we focus on two key model parameters - rank thresholding, i.e., ``crowd size,'' and experience thresholding.  In addition, we develop intuition and provide context through a series of representative models drawn from the model space.
	
	\subsection{Model Test Space}
	To construct our model test space, we select a range of model configurations and hyperparameters.  Below, we describe the configurations and parameters of the model space, beginning with the crowd subset rules and parameters tested.
	
	\begin{enumerate}
		\item \textbf{Experience Threshold}: We evaluate crowd models with and without experience thresholding as defined in Section \ref{section:crowd_subsets} above.   For models with 
		      experience thresholding, we evaluate thresholds from $1$ to $99$, inclusive, by steps of $1$.  In figures and tables below, we represent models without experience thresholding as having a threshold of $0$.
		      
		\item \textbf{Performance Threshold}: We evaluate crowd models with and without performance thresholding as defined in Section \ref{section:crowd_subsets} above.  For models with performance thresholding, we evaluate accuracy-based thresholds with two cutoffs - 62\%, representing the ``always guess reverse'' null model's historical accuracy rate, and 70\%, representing the approximate accuracy of the purely data-driven models.\cite{katz2017general} 
		      
		\item \textbf{Rank Threshold}: We evaluate crowd models with and without rank thresholding as defined in Section \ref{section:crowd_subsets} above.  For models with rank thresholding, we evaluate two types of rank thresholds: one based on experience and one based on accuracy.  For both rank threshold types,  we evaluate rank threshold ranges from $1$ to $99$, inclusive, by steps of $1$.	
		      
		\item \textbf{Statistical Threshold}: We evaluate crowd models with and without statistical thresholding as defined in Section \ref{section:crowd_subsets} above.   For models with statistical thresholding, we evaluate using a one-sided binomial test with $p$-value thresholds.  For each participant in each case, we perform the test using all of their prior case predictions to determine whether the Bernoulli probability of their accuracy distribution exceeds the null model.  If the $p$-value of this test is strictly less than our threshold, then the participant is included in our crowd subset.  We test three $p$-value thresholds - $0.01$, $0.05$, as common modern statistical thresholds, and $0.5$, as a decision-theoretic threshold.
		      
		      Next, we describe the configurations and parameters used for our linear crowd models $M^j(d)$.  In this paper, no ``machine learning'' crowd models as described in Section \ref{section:machine_learning_models} are presented, although some literature refers to exponential weighting as such.
		      
		\item \textbf{Linear Crowd Models}: We evaluate linear crowd models with three types of weights: equal weight, linear weight, and exponential weight.  For linear and exponential weight models, we test using two factors to determine rank: experience and accuracy.  For exponential weight models, we test with three values of $\alpha$: $1.0$, $0.5$, and $0.1$.
	\end{enumerate}
	
	In total, this combination of crowd subset and weighting rules and parameters produces a model space with 277,201 models.  This number corresponds to $1 + 99 \cdot 100 \cdot 28$, where the first term corresponds to the simplest crowdsourcing model with no subset or weighting rules, and the second term corresponds to 28 models - each combination of threshold or weighting rules - for each combination of 99 rank and 100 experience thresholds.  We exhaustively sample this space using empirical prediction data from FantasySCOTUS as described in Section \ref{section:data} above.  While we highlight a set of general and specific results, we direct the interested reader to this publication's website, where source code, data, and results will be available.
	
	\subsection{Results}
	For each of our 277,201 model configurations, we record information about the model's performance for each Justice in each case, including the number of participants and model score for the crowd.  Using this information, we can assess the performance of these models in both nuanced or aggregate ways.  Doing so allows us to address questions like whether crowds better predict conservative or liberal Justices or whether crowds perform better at the beginning or end of each term. 
	
	\subsubsection{Baseline}
	Within the context of this research, we focus on the aggregate performance of the crowd across all cases.  However, performance is not just a simple calculation of accuracy; instead, performance assessment should instead capture a model's strengths and weaknesses relative to existing prediction methods.  To accomplish this goal, we evaluate both accuracy and F1 scores relative to a fair ``baseline'' or ``null'' model.  In prior related work, we highlight some commonly held wisdom regarding the appropriate null model in this context:
	
	\begin{quotation}
		[...] few legal experts would rely on an unweighted coin as a null model against which to compare their predictions.  Instead, informed by recent years, common wisdom among the legal community is that the baseline betting strategy should be to \textit{always guess Reverse}.  This strategy is supported by the recent history of the Court  over the last 35 terms: 57\% of Justice votes and 63\% of case outcomes have been Reverse.\cite{katz2017general}
	\end{quotation}
	
	Following this commonly-held understanding, we use the ``always guess reverse'' null model in this paper as a baseline to consider both the Justice-level and case-level outcomes.  Within our sample, the proportions are 64.6\% and 61.5\% reverse for case and Justice outcomes, respectively.  From a legal perspective, the Supreme Court usually ``takes a case'' to correct an error below, not to affirm it.
	
	\subsubsection{Example Model Performance}
	It is difficult to understand a forest without first examining a tree or two.  Before presenting the aggregate results of hundreds of thousands of models, we first present four example models so that the reader can develop more understanding about the behavior of individual models.  Below, we describe four models of increasing complexity selected from the model space above.
	
	\begin{enumerate}
		\item \textbf{Simple majority rule}: For a given docket $d$ and a set of participants $S(d)$, let $M^j(d)$ be the simple majority rule voting model.  In other words, $M^j(d) = \frac{1}{|S(d)|} \sum_{i \in S(d)} c^j_i(d)$, i.e., an equal-weighting of all participant predictions.
		      
		\item \textbf{``Follow the Leader(s)'' without experience thresholding:}  For a given docket $d$ and a set of participants $S(d)$, let $M^j(d)$ be the ``Follow the Leader'' (FTL) model without experience thresholding.  In other words, let the crowd subset $S^*$ be defined as $\{i : s(i, \mathcal{I}(T(d)) = 1 \ \forall i \in S(d)\}$, where $s(i, \mathcal{I}(d)) = \mathbb{I}_{R_P(i, \mathcal{I}(d)) = 1}$, and let $M^j(d)$  be an equal weighting or simple majority over this crowd subset.  In cases where there are ties in accuracy, which do occur,  then there are multiple Leaders.
		      
		\item \textbf{``Follow the Leader(s)'' with experience thresholding:} For a given docket $d$ and a set of participants $S(d)$, let $M^j(d)$ be a ``Follow the Leader'' (FTL) model with an experience thresholding requiring at least five cases.  In other words, let the crowd subset $S^*(d)$ be defined sequentially through two filters: first, $S^*(d)$ is calculated as $\{i : s_1(i, \mathcal{I}(T(d)) = 1 \ \forall i \in S(d)\}$, where $s_1(i, \mathcal{I}(d)) = \mathbb{I}_{\text{XP}(i, \mathcal{I}(d)) \geq 5}$; second, $S^*(d)$ is updated as $\{i : s_2(i, \mathcal{I}(T(d)) = 1 \ \forall i \in S^*(d)\}$, where $s_2(i, \mathcal{I}(d)) = \mathbb{I}_{R_P(i, \mathcal{I}(d)) = 1}$. Finally, let $M^j(d)$ be an equal weighting or simple majority over this crowd subset.  In cases where there are ties in accuracy, which do occur, then there are multiple Leaders.
		      
		\item \textbf{Maximum accuracy model}: Finally, to demonstrate the performance ``ceiling'' of 80.8\%, we present the model with the maximum accuracy in our sample.  This model is defined for a given docket $d$ and a set of participants $S(d)$ is defined sequentially through two crowd subset filters.  First, $S^*(d)$ is calculated as $\{i : s_1(i, \mathcal{I}(T(d)) = 1 \ \forall i \in S(d)\}$, where $s_1(i, \mathcal{I}(d)) = \mathbb{I}_{\text{XP}(i, \mathcal{I}(d)) \geq 5}$. Second, $S^*(d)$ is updated as $\{i : s_2(i, \mathcal{I}(T(d)) = 1 \ \forall i \in S^*(d)\}$, where $s_2(i, \mathcal{I}(d)) = \mathbb{I}_{R_P(i, \mathcal{I}(d)) \leq 22}$.  We then apply exponential weighting with an $\alpha$ of $0.1$, i.e., we let $M^j(d)$ be $\frac{1}{\sum_{i \in S^*(d)} \exp(-0.1 \cdot R_P(i))} \sum_{i \in S^*(d)} \exp(-0.1 \cdot R_P(i)) \cdot c^j_i(d)$.
	\end{enumerate}
	
	Table \ref{table:table2} displays both the case-level accuracy and F1 score for each of the Example Models listed above, as well as the \textit{always guess reverse} null model for purposes of baseline comparison.  Table \ref{table:table2} demonstrates that all of these crowdsourcing models produce more accurate results than the null model.
	
	\setlength{\tabcolsep}{0.4em} 
	{\renewcommand{\arraystretch}{1.1}%
		\begin{table}[htb]
			\footnotesize
			\begin{tabular}{|l|r|r|r|r|}
				\hline
				\textbf{Description}                    & \textbf{Accuracy} & \textbf{F1} \\\hline
				Baseline - Always guess reverse         & 64.6              & 78.4        \\
				1 - Simple majority rule                & 66.4              & 69.2        \\
				2 - FTL without experience thresholding & 65.7              & 71.2        \\
				3 - FTL with experience thresholding    & 72.8              & 78.7        \\
				4 - Maximum accuracy                    & 80.8              & 85.2        \\
				\hline
			\end{tabular}
			\caption[]{Case-level performance metrics for example and baseline models}
			\label{table:table2}
		\end{table}
		
		Figure \ref{fig:Fig1} displays the time series of case-level cumulative accuracy for each of the illustrative model configurations. The null model is also included for purposes of comparison.  After some initial volatility, the performance is very consistent over the six-year period of study.  
		
		\begin{figure}[htb]
			\centering
			\includegraphics[width=.5\textwidth]{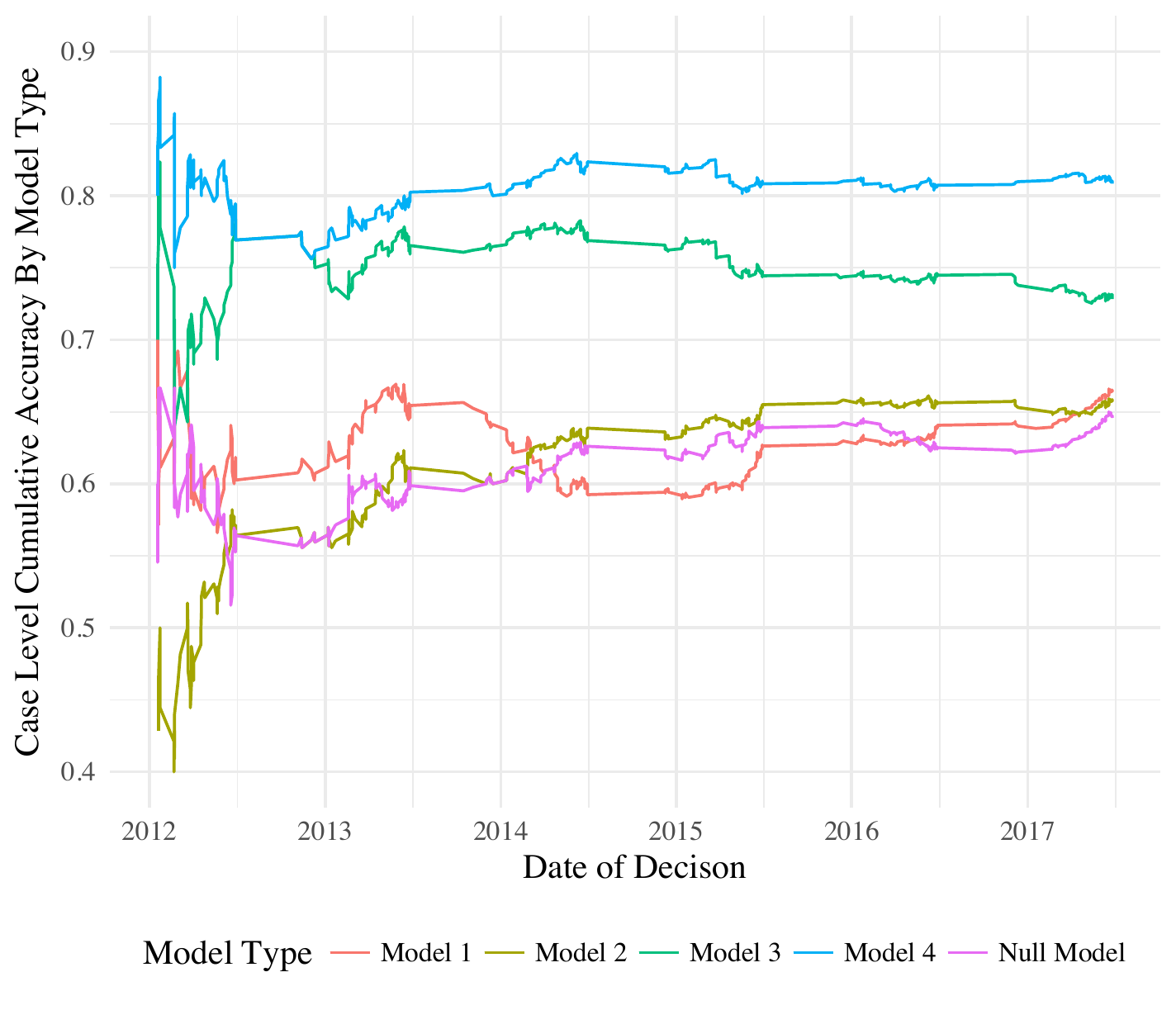}
			\caption{Cumulative case-level accuracy of example and baseline models}
			\label{fig:Fig1}
		\end{figure} 
		
		In addition to the case-level performance, we also display overall Justice-level results, which are slightly lower from an accuracy standpoint but otherwise qualitatively similar.  Table \ref{table:table3} displays the aggregate performance of our example model at the Justice level for the same set of model configurations as Table \ref{table:table2}.
		
		\setlength{\tabcolsep}{0.4em} 
		{\renewcommand{\arraystretch}{1.1}%
			\begin{table}[htb]
				\footnotesize
				\begin{tabular}{|l|r|r|r|r|}
					\hline
					\textbf{Description}                    & \textbf{Accuracy} & \textbf{F1} \\\hline
					Baseline - Always guess reverse         & 61.5              & 76.2        \\
					1 - Simple majority rule                & 65.5              & 71.2        \\
					2 - FTL without experience thresholding & 63.8              & 70.5        \\
					3 - FTL with experience thresholding    & 72.2              & 78.7        \\
					4 - Maximum accuracy                    & 77.1              & 83.1        \\
					\hline
				\end{tabular}
				\caption[]{Justice-level performance metrics for example and baseline models}
				\label{table:table3}
			\end{table}
			
			Figure \ref{fig:Fig2} displays the time series of Justice-level cumulative accuracy for each of the example model configurations. As before, the null model is also included as a baseline for purposes of comparison.  
			
			\begin{figure}[htb]
				\centering
				\includegraphics[width=.50\textwidth]{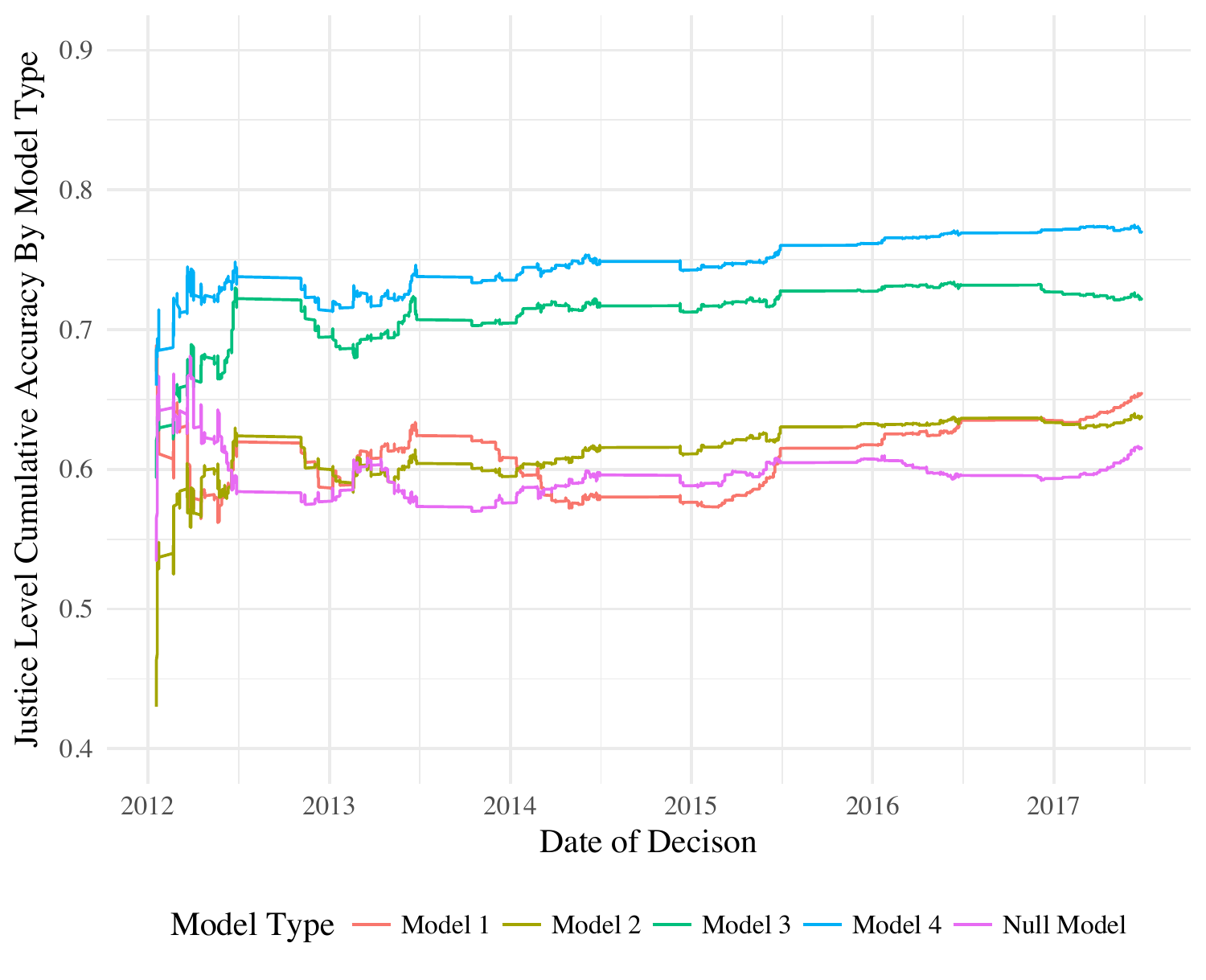}
				\caption{Cumulative Justice-level accuracy of example and baseline models}
				\label{fig:Fig2}
			\end{figure} 
			
			\subsubsection{Model Space Performance}
			In this paper, our goal is not just to demonstrate that a well-selected model such as the four identified above can outperform the null model; rather, our goal is to demonstrate that crowdsourcing, across a wide range of conditions and choices, can outperform other commonly used prediction methods.  To do so, we present in this section both the aggregate accuracy and overall robustness of our model space.
			
			First, we evaluate a number of population-level calculations about our model space.  Across all models, the mean and median Justice-level model accuracy exceeds the null model by 12.0\%.  Similarly, the mean and median Case-level model accuracy exceeds the null model by 9.5\% and 9.6\%, respectively.  All 277,201 (100\%) of models exceed the null model for Justice-level prediction; however, 8 of 277,201 models fail to exceed the null for Case-level predictions.  These 8 models are all characterized by their rank threshold value of 2 and majority-rule vote model, resulting in an inability to break ties in many cases.  More impressively, 86.9\% of models exceed the Justice-level null model accuracy by at least 10\%, and 41.4\% of models exceed the Case-level null model accuracy by at least 10\%.
			
			Second, as noted earlier, we project the model space onto two key dimensions - rank threshold and experience threshold.  These two dimensions are the most commonly used crowd construction parameters.  Rank thresholding is generally used to control the crowd size, allowing modelers to capture either ``Follow the Leader'' behavior with a rank threshold of $1$ or the ``wisdom of the crowd'' as the rank threshold exceeds the crowd size.  Experience thresholding is generally used to control the quality of data about participants and ensure engagement; by setting higher experience thresholds, models have more data to use to make decisions about participants.  
			
			\begin{figure}[htb]
				\centering
				\includegraphics[width=0.53\textwidth]{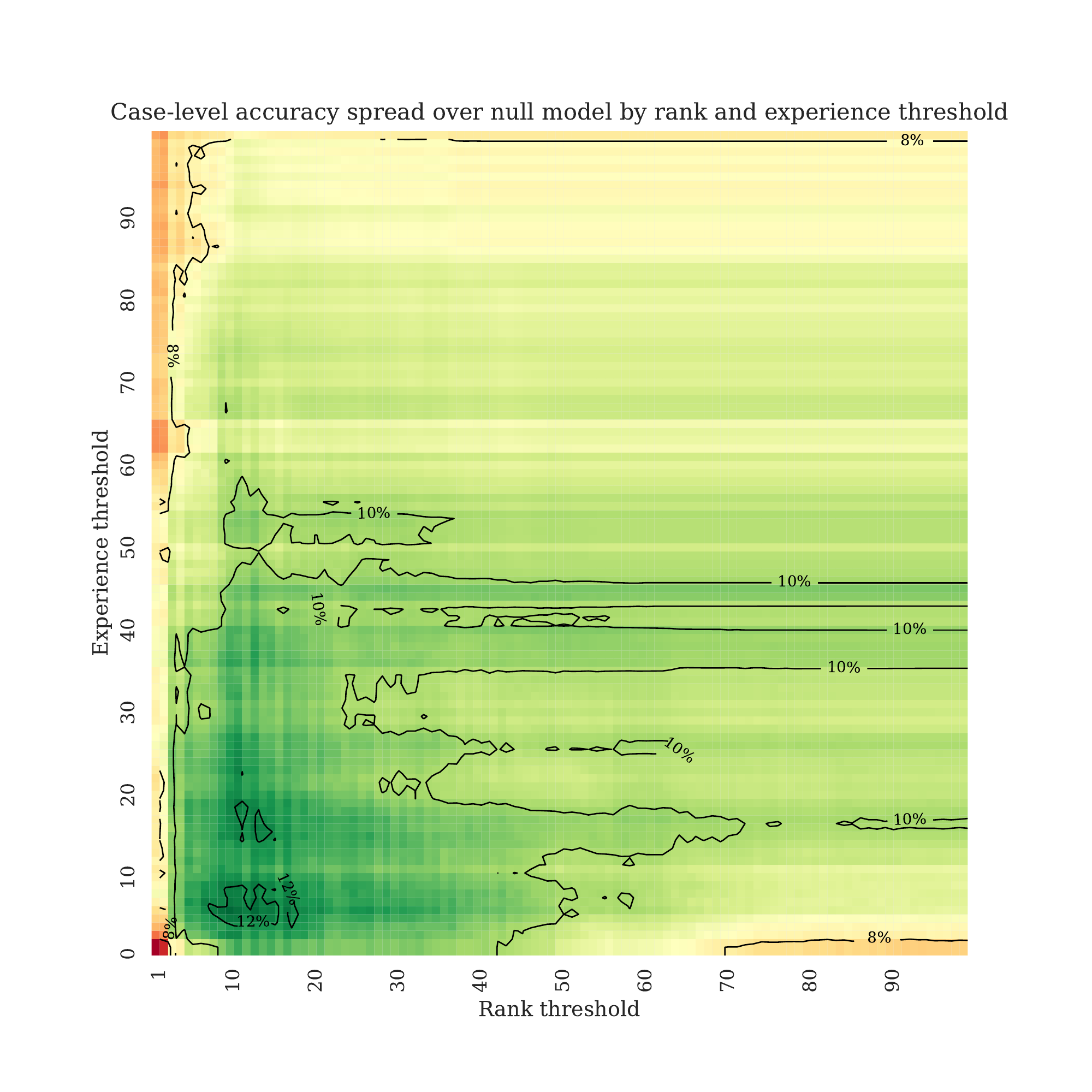}
				\caption{Case-level accuracy spread by rank and experience threshold}
				\label{fig:contour_case_spread}
			\end{figure}
			
			\begin{figure}[htb]
				\centering
				\includegraphics[width=0.53\textwidth]{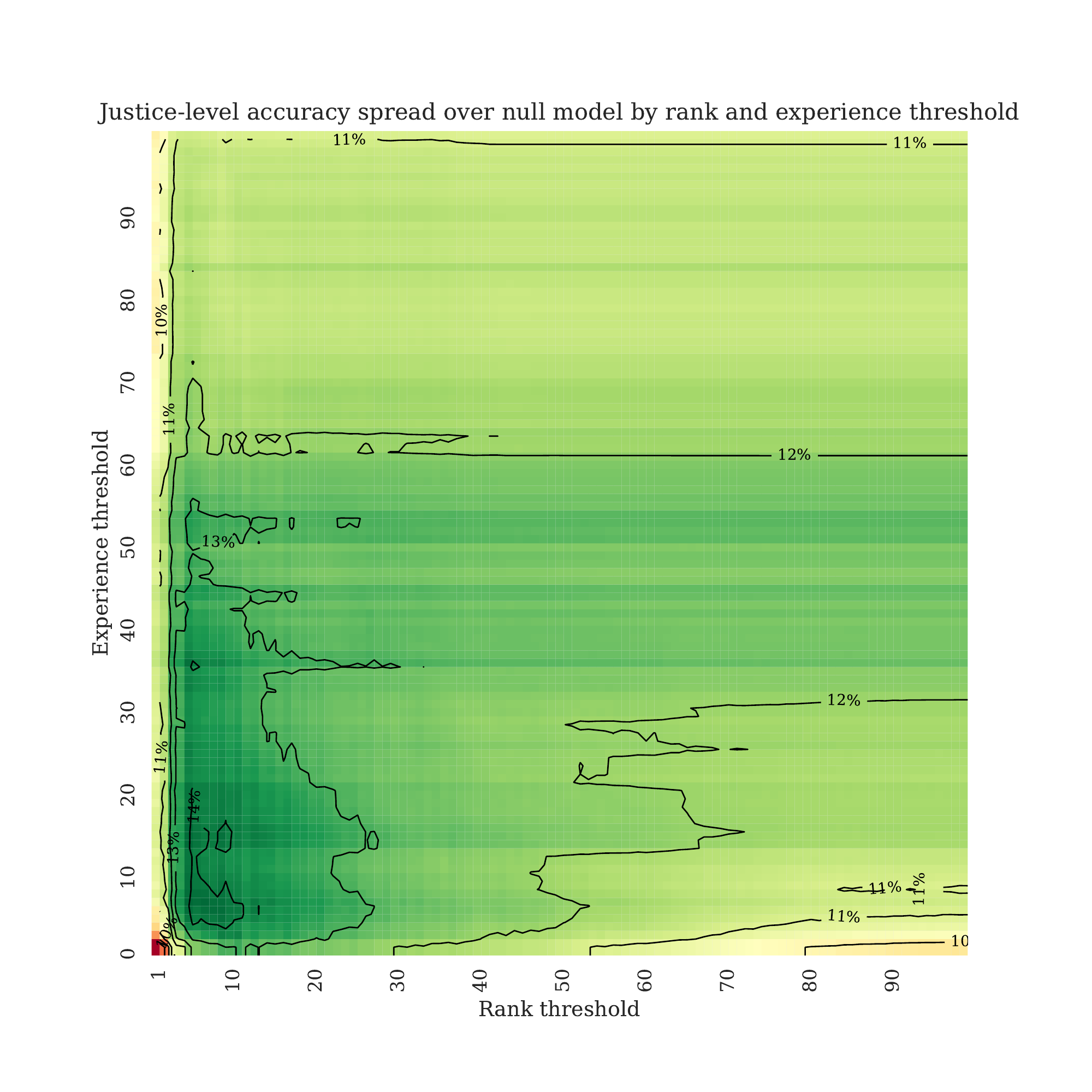}
				\caption{Justice-level accuracy spread by rank and experience threshold}
				\label{fig:contour_justice_spread}
			\end{figure}
			
			In our experiment, we evaluate values of rank thresholds from $1$ to $99$ and values of experience thresholds from $0$ to $99$.  In Figures \ref{fig:contour_case_spread} and \ref{fig:contour_justice_spread}, we present accuracy metrics across a $99 \times 100$ grid, where each cell within the contour plot is an average across 28 models varying all other model construction configurations and parameters.  Figures \ref{fig:contour_case_spread} and \ref{fig:contour_justice_spread} demonstrate that crowdsourcing in this context is both accurate and robust.  
			
			While there is variance across the space, especially visible in the ``corners'' of the parameter space, the vast majority of the contour plots are green, and all 9900 cells have positive averages.  Furthermore, these two figures demonstrate vertical even-odd ``waves,'' hinting at the effect of tie-breaking on even-sized crowds for simpler model types.  Overall, however, both Justice and Case-level parameter spaces are characterized by large, stable performance ``plateaus'' near the origin.  These plateaus are easy to guess; for example, beginning with a crowd size of 10 participants and an experience threshold of 10 cases immediately lands a modeler in the plateau.  Other than the Follow-the-Leader models at the origin, as one moves away from this plateau, performance gradually declines but remains positive relative to the baseline.  Qualitatively similar results are found for F1 scores, which can be examined on this paper's GitHub repository.
				
			\begin{figure}[htb]
				\centering
				\includegraphics[width=0.49\textwidth]{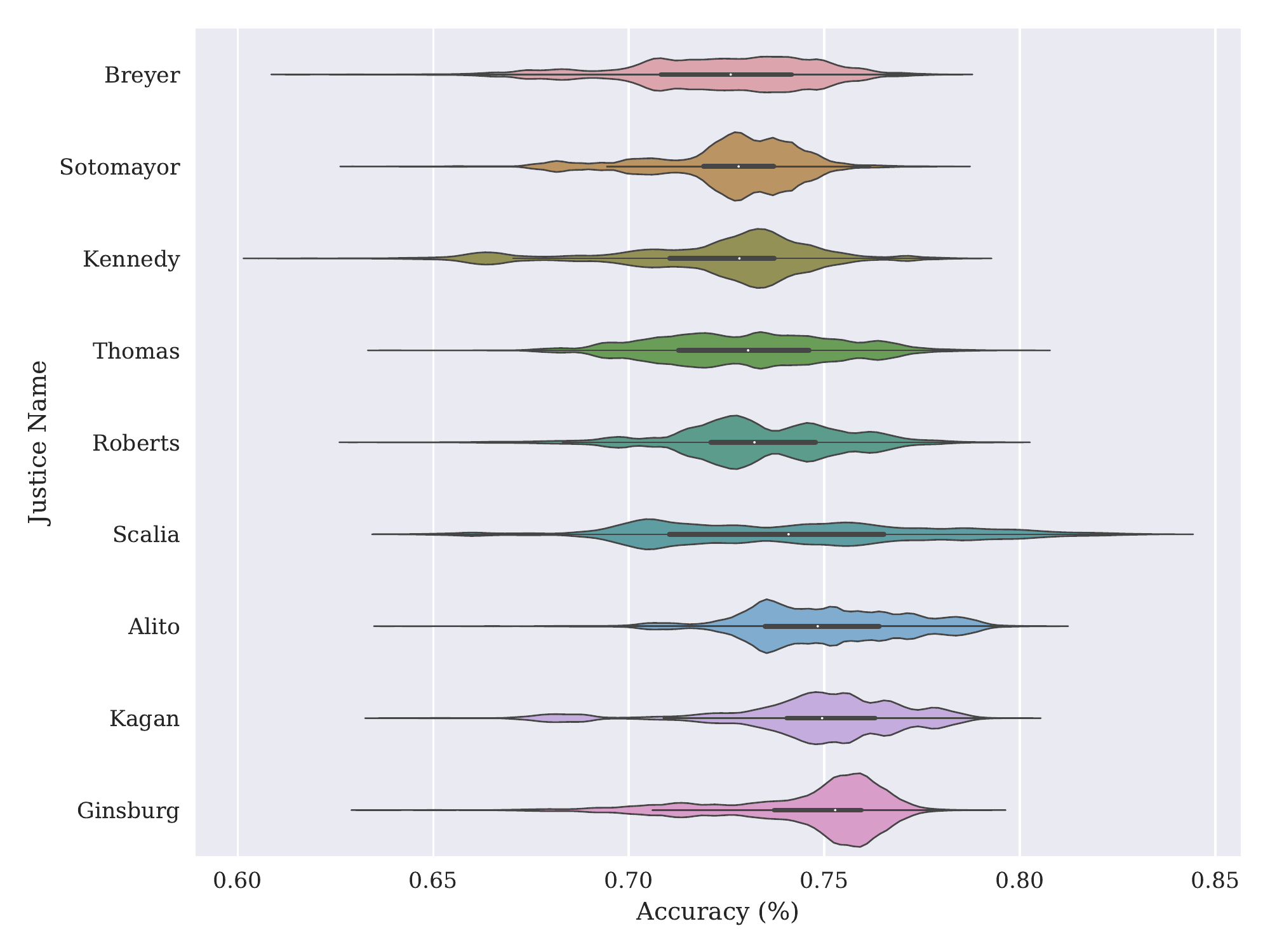}
				\caption{Distribution of model accuracy for individual Justices}
				\label{fig:violin_justice_accuracy}
			\end{figure}
            
            \setlength{\tabcolsep}{0.4em} 
		{\renewcommand{\arraystretch}{1.1}%
			\begin{table}[htb]
				\footnotesize
				\begin{tabular}{|l|r|r|r|}
					\hline
					\textbf{Justice}                    & \textbf{Accuracy (\%)} & \textbf{Baseline (\%)} & \textbf{Sample Size} \\\hline
					Alito        &  75.0 &     61.3 &        445 \\
                    Breyer       &  72.3 &     61.5 &        449 \\
                    Ginsburg     &  74.5 &     61.5 &        449 \\
                    Gorsuch      &  77.4 &     57.1 &         14 \\
                    Kagan        &  74.7 &     61.9 &        443 \\
                    Kennedy      &  72.2 &     63.7 &        449 \\
                    Roberts      &  73.3 &     64.5 &        448 \\
                    Scalia       &  74.0 &     59.8 &        328 \\
                    Sotomayor    &  72.6 &     61.6 &        445 \\
                    Thomas       &  73.0 &     57.5 &        449 \\
					\hline
				\end{tabular}
				\caption[]{Average Accuracy by Justice across all models}
				\label{table:table4}
			\end{table}   
    
	In order to provide additional evidence for the robustness of crowdsourcing, we also examine how crowdsourcing performs for each Justice on the Court.  In Figure \ref{fig:violin_justice_accuracy}, we present a violin plot for the model accuracy across the entire model space for each Justice excluding Gorsuch, who is omitted from the figure due to his very small sample size.  First, we observe a relatively small range of central tendency between Justices; the median model accuracy for all Justices is between 71\% and 76\%.  Second, we observe that while central tendency is relatively tight, the tails and skewness of these Justice-level distributions vary more so.  For example, Breyer, Scalia, and Ginsburg all exhibit large negative and positive tails relative to other Justices.  Third, despite the wide tails for some Justices, we find that the performance for most models for most Justices is well above baseline rates.  Even the minimum accuracy model is greater than the baseline for all Justices other than Breyer, Kennedy, and Roberts, and the $1^{st}$ percentile of model accuracies is greater than the baseline rate for all Justices.  Table \ref{table:table4} provides the average accuracy across all 277,201 models, the baseline ``always guess reverse'' rate, and the sample size for each Justice, including Gorsuch.  These Justice-level findings provide additional support for the accuracy and robustness of crowdsourcing in this context. 
    	
			\section{Conclusion}
			\label{section:conclusion_future_work}
            In this paper, we provide strong support for the claim that crowdsourcing can accurately and robustly predict the decisions of the Supreme Court of the United States.  More importantly, unlike most extant literature on crowdsourcing, we do so through the application of empirical model thinking.  This approach allows us to confidently demonstrate that, across a wide range of choices regarding crowdsourcing model configurations and parameters, crowdsourcing outperforms both the commonly accepted ``always guess reverse'' model and the best-studied algorithmic models.
            
            In the process of supporting these results, we also produce two additional contributions - an open crowdsourcing dataset and a comprehensive, formal framework for crowdsourcing model construction.  To our knowledge, this dataset represents one of the largest available sources of recurring human prediction to date, and we encourage other researchers to build on this resource in future research.  Furthermore, we believe our model construction framework can help researchers design and communicate more effectively across a range of theoretical and applied crowdsourcing problems.
            
            In future work, we will examine the application of more sophisticated ``machine learning'' models, as discussed in Section \ref{section:crowd_construction_framework} above, as well as how information such as subject matter, plaintiff, defendant, or Justice can be used to improve the performance of Supreme Court prediction models.  Lastly, we intend to apply our empirical model thinking to other crowdsourcing contexts, such as the prediction of elections or legislative outcomes.

			\bibliography{crowdsourcing.bib}
			
			
\end{document}